\documentclass[12pt]{article}
\usepackage{amssymb,amsmath,amsthm,graphicx,ulem}

\pdfoutput=1
\usepackage{hyperref}
\usepackage{graphicx,subfigure}
\usepackage{epsfig}
\usepackage{amsmath}
\usepackage{amsfonts}
\usepackage{amssymb}
\usepackage[usenames]{color}
\usepackage[letterpaper,left=2.cm,right=2.cm,top=2.5cm,bottom=2.5cm]{geometry}

\newcommand{\beq}{\begin{equation}}
\newcommand{\eeq}{\end{equation}}
\newcommand{\be}{\begin{equation}}
\newcommand{\ee}{\end{equation}}
\newcommand{\beqa}{\begin{eqnarray}}
\newcommand{\eeqa}{\end{eqnarray}}
\newcommand{\beqar}{\begin{eqnarray*}}
\newcommand{\eeqar}{\end{eqnarray*}}
\newcommand{\bea}{\begin{eqnarray}}
\newcommand{\eea}{\end{eqnarray}}

\newcommand{\p}{\partial}

\newcommand{\reef}[1]{(\ref{#1})}

\newcommand{\ie}{{\it i.e.}\ }

\newcommand{\dd}{\textrm{d}}

\def\dd{\mathrm{d}}

\numberwithin{equation}{section}

\newcommand{\nn}\nonumber

\numberwithin{equation}{section}

\begin{document}

\allowdisplaybreaks

\normalem

\begin{center}

{ \Large {\bf Geons and the Instability of Anti-de Sitter Spacetime}}

\vspace{1cm}

Gary T. Horowitz$^{1}$ and Jorge E. Santos$^{2,3}$

\vspace{1cm}

{\small
$^{1}${\it Department of Physics, UCSB, \\
Santa Barbara, CA 93106, USA}}

\vspace{0.5cm}

{\small
$^{2}${\it Department of Physics, Stanford University, \\
Stanford, CA 94305-4060, USA }}

\vspace{0.5cm}

{\small
$^{3}${\it Department of Applied Mathematics and Theoretical Physics, \\
University of Cambridge, Wilberforce Road, \\
Cambridge CB3 0WA, UK}}

\vspace{1.6cm}

\end{center}

\begin{abstract}
\noindent  We briefly review the evidence that anti-de Sitter spacetime is nonlinearly unstable, and the perturbative arguments that there should exist geons - nonsingular solutions to Einstein's equation with a helical symmetry.  We then explicitly construct these geons numerically and discuss some of their properties. We conclude with some open questions. 
 \end{abstract}

\newpage


\tableofcontents
\baselineskip16pt

\section{Introduction}

Anti-de Sitter spacetime (AdS) is the Lorentzian space with constant negative curvature. In four spacetime dimensions (which will be the focus of our discussion) the metric can be written
\begin{equation}\label{AdS}
\mathrm{d}s^2 = -\left(1+\frac{r^2}{L^2}\right)\mathrm{d}t^2+\left(1+\frac{r^2}{L^2}\right)^{-1}{\mathrm{d}r^2}+r^2(\mathrm{d}\theta^2+\sin^2\theta \mathrm{d}\phi^2)
\end{equation}
where $L$ sets the scale of the curvature and is called the AdS radius. This is the maximally symmetric solution to Einstein's equation with a negative cosmological constant:
\be\label{EE} 
R_{ab} = -{3\over L^2} g_{ab}
\ee
Its isometry group is $SO(3,2)$. For most of the hundred year history of general relativity, this solution was just a curiosity that attracted little interest. However, due to developments in string theory, there has recently been an explosion of interest in spacetimes that approach AdS at infinity. This is because of the discovery of gauge/gravity duality which states that solutions of \reef{EE} that asymptotically approach AdS can be used to understand properties of certain strongly coupled quantum field theories.\footnote{More generally, the duality states that a theory of quantum gravity with AdS boundary conditions is equivalent to a quantum field theory (without gravity). We will be interested in just the classical gravity limit of this correspondence.}

AdS is not globally hyperbolic. If one conformally rescales the metric \reef{AdS} by $L^2/r^2$, the spacetime is mapped to the interior of a finite cylinder with boundary $S^2 \times R$  and metric 
\be\label{bdymetric}
\mathrm{d}s^2_{bdy} =  -\mathrm{d}t^2 + L^2(\mathrm{d}\theta^2+\sin^2\theta \mathrm{d}\phi^2) 
\ee
To evolve asymptotically AdS initial data, one needs to specify boundary conditions at infinity. Friedrich \cite{Friedrich:1995vb} has shown that Einstein's equation has a well posed initial value problem for various choices of boundary conditions. The standard one adopted by gauge/gravity duality is to fix the boundary metric.\footnote{Actually, it is just the conformal class that needs to be fixed.} The simplest choice, and the one we will adopt, is to take this metric  to be \reef{bdymetric}. Throughout the text, greek indices will run over the boundary directions.

For asymptotically flat spacetimes, one can define a total  energy-momentum vector at spacelike or null infinity. Since the conformal boundary of an asymptotically AdS spacetime is timelike, one can define a boundary stress energy tensor $T_{\mu\nu}$ from which one can compute conserved quantities associated with asymptotic symmetries\footnote{This should not be confused with the usual stress energy tensor of matter coupled to gravity. This boundary stress energy tensor can be nontrivial even for pure gravity solutions satisfying \reef{EE}.}  \cite{Balasubramanian:1999re,deHaro:2000xn}. The best way to define $T_{\mu\nu}$ is to introduce a set of coordinates that is well adapted to the expansion of the Einstein equations off the conformal boundary. Such coordinates were first introduced by Fefferman and Graham in their seminal paper \cite{0602.53007}, and take the following form
\begin{equation}
\mathrm{d}s^2 = \frac{L^2}{z^2}\Big[\mathrm{d}z^2+g_{\mu\nu}(z)dx^\mu dx^\nu \Big]\,,
\end{equation}
where the boundary is now located at $z=0$. In \cite{0602.53007}, the authors have further shown that Einstein's equations ensure that $g_{\mu\nu}(z)$ has a well defined power series expansion:\footnote{This provides a definition of asymptotically AdS spacetimes.}
\be
g_{\mu\nu}(z) = g_{\mu\nu}^{(0)} + g_{\mu\nu}^{(1)}z + g_{\mu\nu}^{(2)}z^2 + g_{\mu\nu}^{(3)}z^3 + \cdots
\ee
In this expansion, the first term on the right is an arbitrary Lorenzian metric on the boundary and the next two terms are completely fixed in terms of $g_{\mu\nu}^{(0)}$. The following term, $g_{\mu\nu}^{(3)}$, is not fixed by the asymptotic equations, and depends on the solution in the interior. The asymptotic stress tensor $T_{\mu\nu}$ is simply proportional to  $g_{\mu\nu}^{(3)}$.

A more covariant definition of  $T_{\mu\nu}$ can be given as follows. Fix a small $z_0$ and consider the timelike surface $\partial \mathcal{M}_{z_0}$ at  $z = z_0$. Let  $\gamma_{\mu\nu}$ be the induced metric, ${\cal R}_{\mu\nu}$ its Ricci tensor, and $K_{\mu\nu}$  the extrinsic curvature of this surface (with respect to the inward pointing normal).
Then we set
\be
\tilde T_{\mu\nu} =  \frac{1}{8\pi G}\left [ K_{\mu\nu} - K \gamma_{\mu\nu} - \frac{2}{L} \gamma_{\mu\nu} + L ({\cal R}_{\mu\nu} - \frac{1}{2} {\cal R} \gamma_{\mu\nu}) \right ]
\ee
Einstein's equations ensure that for any $z_0$, this stress tensor is conserved, $D_\mu \tilde T^{\mu\nu} = 0$, where $D_\mu$ is the covariant derivative  with respect to $\gamma_{\mu\nu}$. (The last two terms are the Einstein tensor of $\gamma_{\mu\nu}$  which is automatically conserved, and  the conservation of the first two terms  is essentially the momentum constraint equation for evolution in the $z$ direction.)
The boundary stress tensor is simply 
\be\label{bdystressenergy}
 T_{\mu\nu} = \lim_{z_0 \to 0} \tilde T_{\mu\nu} .
 \ee

When $g_{\mu\nu}^{(0)}$ is given by \reef{bdymetric},
an asymptotically AdS spacetime has a large number of asymptotic Killing vectors. We will mainly be interested in two: given the asymptotic timelike Killing vector 
$\xi = \p_t$ and any spacelike cross-section $S$ of $\partial \mathcal{M}_{z_0}$, the total energy (with respect to $\xi$) is 
\be\label{energy}
E = \lim_{z_0 \to 0} \int_S \tilde T_{\mu\nu}\xi^\mu n^\nu dS
\ee
where $n^\nu$ is the unit timelike normal to $S$ in $\partial \mathcal{M}_{z_0}$. Similarly, given the asymptotic rotational Killing vector $\eta = \p_\phi$, the total angular momentum is 
\be\label{angmomentum}
J = \lim_{z_0 \to 0} \int_S \tilde T_{\mu\nu}\eta^\mu n^\nu dS
\ee

Since null geodesics in \reef{AdS} reach infinity in finite time,  waves of massless fields propagating in the spacetime can reach infinity in finite time. Since energy is conserved,  these waves are reflected back to the interior. In effect,  the spacetime  acts like a finite box. For this reason, the boundary condition we have adopted is sometimes called the reflecting boundary condition.

A key question is whether AdS with this boundary condition is nonlinearly stable. The nonlinear stability of Minkowski spacetime and de Sitter spacetime (having constant positive curvature) was established some time ago \cite{Christodoulou:1993uv,Friedrich86}. Given the current interest in gauge/gravity duality, it is important to determine whether a similar result is true for AdS.  It is well known that AdS is  stable at the linearized level. There is an infinite tower of normal modes with frequencies that are all integer multiples of $1/L$. At the nonlinear level there are at least two reasons to suspect things might be different. Physically, since AdS is like a confining box, if Einstein's equations were sufficiently ergodic, any energy added to it would explore all possible configurations of the given energy, which include small black holes. Mathematically, since the linearized modes do not decay, nonlinear corrections (which typically involve integrals of powers of the linearized modes) can become large.

There has been recent evidence that AdS is indeed nonlinearly unstable. It appears that generically, small finite perturbations evolve to form black holes. This was first seen in a model of gravity coupled to a spherically symmetric massless scalar field  \cite{Bizon:2011gg}.
In this case, the metric has no independent degrees of freedom and is completely determined by the scalar field. Given an initial profile $\phi = A \psi(r)$ for the scalar field, one can solve the Einstein constraint equations to determine the initial metric, and then numerically evolve to determine the solution for all later time. For fixed profile $\psi(r)$, and large amplitude $A$, the scalar field contains a lot of energy and it collapses down to form a large black hole. As one decreases $A$, the black hole that is initially formed gets smaller and at a critical amplitude $A_1$, one forms a ``zero mass black hole" which is really a naked singularity. This was first found in studies of spherically symmetric scalar field collapse in asymptotically flat spacetimes \cite{Choptuik:1992jv,Christodoulou:1991yfa}. In that case, initial scalar fields with $A < A_1$ just scatter and go off to infinity leaving a globally nonsingular spacetime. In AdS, these waves reflect off infinity and return to the interior. It was found that for $A$ slightly smaller than $A_1$ these reflected waves form a black hole. As one continues to decrease the amplitude, the black hole formed by the reflected waves gets smaller and there is a second critical amplitude $A_2$ where one again forms a naked singularity.
For $A$ slightly smaller than $A_2$, the scalar field reflects off infinity twice and then forms a black hole. This
pattern appears to continue indefinitely. The numerical evidence suggests that no matter how small the initial amplitude is, one eventually forms a black hole. One finds that for small amplitude, the time to form a black hole scales like $1/A^2$.

There is also an analytic perturbative explanation for this instability based on resonances. Since the linearized modes all have equal spaced frequencies, when one computes higher order corrections, there are typically secular terms which grow with time. Some of these can be removed by adjusting the frequencies of the linear modes (effectively resuming part of the perturbation expansion), but others cannot. It was argued \cite{Bizon:2011gg} that it was the growth of these secular terms which led to the breakdown of perturbation theory and the nonlinear instability. While this is the generic behavior, it was also found that if one starts with a single mode of the scalar field, there appeared to be no problem adding higher order corrections and keeping the solution time periodic. Indeed these ``oscillons" have since been constructed numerically \cite{Maliborski:2013jca}. They represent exact, time periodic, spherically symmetric solutions of the Einstein-scalar field equations.

Since this model is spherically symmetric, no gravitational degrees of freedom are excited. We now investigate whether pure gravity, without any assumptions of symmetry, behaves in the same way. The following section is based on  \cite{Dias:2011ss}.

\section{Perturbative analysis}

To find perturbative solutions to \reef{EE}, one expands the metric about the AdS background, \emph{i.e.} $g = \bar{g}+\sum_i \epsilon^i h^{(i)}$, where $\epsilon$ is  a perturbation parameter whose physical meaning will be discussed later, and $\bar{g}$ is the metric of AdS \reef{AdS}.
At each order in perturbation theory, the Einstein equations yield
\begin{equation}
\Delta_L h_{ab}^{(i)} = T^{(i)}_{ab},
\label{eq:perturb}
\end{equation}
where $T_{ab}^{(i)}$ is a function of $\{h_{ab}^{(j\leq i-1)}\}$ and their derivatives and $\Delta_L$ is a second order operator constructed from $\bar{g}$,
\begin{equation}
2\Delta_L h_{ab}^{(i)} \equiv -\bar \nabla^2 h_{ab}^{(i)}-2 \bar{R}_{a\phantom{c}b\phantom{d}}^{\phantom{a}c\phantom{b}d}h_{cd}^{(i)}-\bar{\nabla}_{a}\bar{\nabla}_b h^{(i)}+2\bar{\nabla}_{(a} \bar{\nabla}^c h^{(i)}_{b)c}.
\end{equation}
Here, $h^{(i)}\equiv \bar{g}^{ab}h_{ab}^{(i)}$, and $\bar{R}_{abcd}$ is the AdS Riemann tensor. As a consequence of the Bianchi identities, $\bar \nabla^a T^{(i)}_{ab} =0$ for each $i$.

Since global AdS$_4$ contains a round 2-sphere, one can decompose perturbations according to how they transform under diffeomorphisms on the sphere. Normal modes of AdS$_4$ then come in two distinct classes: vector and scalar type gravitational perturbations \cite{Regge:1957td,Edelstein:1970sk,Vishveshwara:1970cc,Zerilli:1970se,Zerilli:1971wd,Ishibashi:2004wx}. Scalar-type gravitational perturbations depend on three integers. Two of these, $\ell_s\geq 2$ and $|m_s|\leq\ell_s$, label the usual scalar spherical harmonics. The remaining integer, $p_s \geq 0$, labels the number of nodes in the radial direction. Likewise, vector-type gravitational perturbations depend on three parameters. Two of these, $\ell_v\geq1$ and $|m_v|\leq\ell_v$, label the usual vector spherical harmonics and $p_v\geq0$, labels the number of nodes in the radial direction.
 
One can  solve the equations \reef{eq:perturb} order by order assuming regularity at the origin and a fall off asymptotically so the metric remains asymptotically AdS. At first order, one finds that scalar-type gravitational perturbations exhibit a harmonic time dependence, $\cos (\omega t)$, with frequencies
 $L\omega= 1+\ell_s +2 p_s$. In \cite{Dias:2011ss} two cases were studied, both corresponding to scalar-type first order gravitational perturbations with $p_s=0$. In one case  the first order perturbation had only the single mode $\ell_s =2, m_s = 2$ and the other was a linear combination of two modes:  $\ell_s =2, m_s = 2$ and $\ell_s =4, m_s = 4$. Note that at higher orders in $\epsilon$, vector-type gravitational perturbations will also get excited.

When the linear solution consists of just a single scalar-type gravitational mode $\ell_s =2, m_s = 2$, its $t,\phi$ dependence takes the form
$\cos (3t/L - 2\phi)$. So it is invariant under a helical Killing vector $\p_t + (3/2L) \p_\phi$. Near the origin, this vector is timelike, but asymptotically it is spacelike near the equator and only timelike near the poles of the two-sphere. One now adds higher order corrections.
At second order there are no resonances, and $h_{ab}^{(2)}$  remains bounded in time. At third order there is one resonant mode, but it can be removed by promoting the frequency of our initial mode to be a function of $\epsilon^2$: $L\,\omega_{2} = 3-14703\,\epsilon^2/17920 $. Thus, to third order, the solution is regular everywhere, with no growing modes in time. It is invariant under a Killing vector which is a slightly shifted version of the symmetry of the linearized mode $K \equiv\partial_t+\frac{\omega_2}{2}\partial_\phi$.  One can compute $T_{ab}^{(4)}$, and calculate the energy as a function of the angular momentum to fifth order in $\epsilon$:
\begin{equation}\label{geonEJ}
E = \frac{3J}{2L}\left(1-\frac{4901\,J}{7560\pi L^2}\right),\quad \omega_2 = \frac{3}{L}\left(1-\frac{4901\,J}{3780\pi L^2}\right),
\end{equation}
where we defined $\epsilon$ by $J = \frac{27}{128}\pi L^2\epsilon^2$. One can extend this solution to fifth order with similar results and we expect that it can be  continued to all orders. This suggests that there exist exact solutions to Einstein's equations describing  {\it geons} -  nonlinear generalizations of the linearized gravitational mode with a helical symmetry.  We will confirm this expectation in the next section by numerically constructing the nonlinear geons. The name geon comes from Wheeler's idea of a localized gravitational excitation held together by its own self gravity.

If our first order  solution consists of a linear combination of two scalar-type gravitational modes with $\ell_s =2, m_s = 2$ and $\ell_s =4, m_s = 4$, the higher order behavior is different. At second order, there are again no resonances, so $h_{ab}^{(2)}$  remains bounded in time. At third order one now finds four potential resonances, but only three actually arise. Two can be removed by adjusting the frequencies of our original two modes,   but the resonant mode with the highest possible frequency, $L\,\omega_{6} = 7,\,m_s = \ell_s = 6$ cannot be 
removed. It leads to a linear growth in time for $h_{ab}^{(3)}$. In this pure gravity case, we do not yet have full numerical evolutions to see what happens at late time. However the structure of the perturbation theory one finds in this case is very similar to what one finds for the spherically symmetric scalar field\footnote{For example, our resonant mode matches the general results of \cite{Craps:2014vaa} for the scalar case.}. In particular, since the linearly growing term first appears at third order, after a time $t \sim 1/ \epsilon^2$ this term will be comparable in size to the leading term and perturbation theory will break down. This agrees nicely with the timescale for black hole formation found in the numerical evolution of the spherically symmetric scalar field. In addition, the fact that the mode with highest frequency is the one that grows, is consistent with the fact that in the spherically symmetric case, the energy is transferred  to higher and higher frequencies in a manner similar to a turbulent cascade \cite{Bizon:2011gg}.

Although we explicitly studied only these two cases, we expect that the results will be similar for other modes. In particular, starting with any individual linearized gravitational mode, one should be able to add higher order corrections and perturbatively construct a geon. Starting with generic superpositions of linearized modes, one should encounter resonances which trigger an instability.

We should point out that there is some controversy surrounding this explanation for the instability of AdS in terms of resonances. Numerical studies of gravity coupled to a spherical scalar field in a finite cavity has led to conflicting results. By changing the boundary conditions for a massless scalar field, or by adding a mass one can remove the resonances in the linear spectrum. Some numerical evolutions show that black holes form only when the linear spectrum has resonances \cite{Maliborski:2014rma}, while other authors \cite{cardoso} find
that collapse always occurs, even when the linear spectrum is not resonant. If this is correct, there must be another explanation for the instability. Another numerical evolution of two mode initial data \cite{Balasubramanian:2014cja} did not find an instability.  One difficulty with numerical evolutions is that they only last a finite time. Since the timescale for the instability is expected to grow like $ 1/ \epsilon^2$, one cannot probe arbitrarily small amplitudes $ \epsilon$.

\section{Numerical construction of geons}

In this section, we go beyond the perturbative construction of geons in the previous section and numerically construct a one parameter family of exact solutions to \reef{EE}. These solutions are all
 smooth horizonless geometries with a single Killing vector field. This Killing field is helical and, at asymptotic infinity, it can be expressed as a linear combination of the generator of time translations $\partial_t$ and rotational symmetry $\partial_\phi$:
\be\label{defK}
K = \partial_\tau = \partial_t+\Omega \partial_\phi\,,
\ee
where $\Omega$ is a real number, that we coin angular velocity, and that we will use to parametrize our numerical solutions.

We begin by choosing a coordinate system that is adapted to $\partial_\tau$. Without loss of generality, our line element can be written in the following form
\begin{multline}
\mathrm{d}s^2 = \frac{L^2}{(1-y^2)^2}\Big[-Q_1\,G_1(y)\,(\dd \tau+Q_6 \dd y)^2+\frac{Q_2\,\dd y^2}{G_2(y)}+\frac{4\,y^2\,Q_3}{2-x^2}\left(\dd x+Q_7\dd \tau+\frac{Q_8}{y} \dd y\right)^2\\
+y^2\,(1-x^2)^2\,Q_4\,\left(\dd \psi+Q_5 \dd \tau+\frac{Q_9\dd x}{1-x^2}+\frac{Q_{10}}{y} \dd y\right)^2\Big]\,,
\label{eq:line}
\end{multline}
where
$$
G_1(y)=1-y^2+y^4\,\quad\text{and}\quad G_2(y)= \frac{G_1(y)}{(1+y^2)^2}\,,
$$
are functions of $y$ only, and $Q_I$ are ten functions of $\psi$, $y$ and $x$ to be determined numerically. Here, $x\in (-1,1)$, $y\in(0,1)$ and $\psi\in(0,2\pi)$. A couple of words are in order regarding our line element. First, if $Q_I=1$ for $I\leq 4$, $Q_I=0$ for $I\geq6$ and $Q_5 = \Omega$, the line element above reduces to the standard line element of AdS$_4$ in global coordinates \reef{AdS}, so long as we identify
\begin{equation}
t = \tau\,,\quad \phi = \psi+\Omega \tau\,,\quad r = \frac{y}{1-y^2}\,\quad\text{and}\quad x\sqrt{2-x^2} = \cos\theta\,.
\end{equation}
From this it is clear that the conformal boundary is located at $y=1$ and that both $\psi$ and $\phi$ have period $2\pi$. Furthermore, $y=0$ is the AdS center and $x=-1,1$ are the south and north poles, respectively.

As discussed in the previous section, geons are continuously connected to pure AdS and can be viewed as nonlinear realizations of normal modes of AdS. We expect the geons associated with scalar-type gravitational perturbations to be uniquely characterized  by four numbers: three integers $(\ell_s,m_s,p_s)$ and a real number $\epsilon$, which we can regard as parametrizing the geon energy or angular momentum.

We will focus on the case discussed in the previous section: scalar-type geon with $\ell_s = m_s =2$ and $p_s=0$. Scalar-type geons that have even $\ell_s$ and $m_s$ have two additional key properties that we will extensively use in their numerical construction. First, they preserve the symmetry $x\to-x$. This allows us to effectively reduce our integration domain in $x$ to $x\in(0,1)$. Also, they respect the symmetry $\psi \to\psi+\pi$, \ie $\psi$ has period $\pi$. Both of these symmetries can be derived if we note the following: scalar and vector type normal modes form a basis for smooth horizonless geometries in AdS$_4$ \cite{Ishibashi:2004wx}. This means that any solution of the form we want to consider here must be solely expressed as a infinite sum of such tensor harmonics. Now, imagine we start with a scalar-type seed solution that at the linear level only contains $m_s=\ell_s=2$. Its parity under the exchange of $x\to-x$ is simply given by $(-1)^{\ell_s}=1$. At second order, vector-type harmonics will show up, but fortunately, they only have odd $\ell_s$. Since the parity of the vector-type harmonics is $(-1)^{\ell_s+1}$, we see that they also preserve the symmetry we mentioned above.

Our integration domain is then reduced to $y\in(0,1)$, $x\in(0,1)$ and $\psi\in(0,\pi)$. In order to solve the Einstein equations
\begin{equation}
G_{ab}\equiv R_{ab}-\frac{3}{L^2}g_{ab}=0\,,
\label{eq:einstein}
\end{equation}
we will use the De-Turck method, which was first introduced in \cite{Headrick:2009pv} and studied in great detail in \cite{Figueras:2011va} for specific gravitational configurations. The DeTurck method is based on the so called harmonic formulation of the Einstein equations, which can be obtained from Eq.~(\ref{eq:einstein}), by adding the following new term
\begin{equation}
G^{H}_{ab} \equiv G_{ab}-\nabla_{(a}\xi_{b)}=0,
\label{eq:einsteindeturck}
\end{equation}
where $\xi^a = g^{cd}[\Gamma^a_{cd}(g)-\bar{\Gamma}^a_{cd}(\bar{g})]$ and $\bar{\Gamma}(\bar{g})$ is the Levi-Civita connection associated with a reference metric $\bar{g}$. The reference metric is chosen to be such that it has the same asymptotics and regularity structure as $g$. For the case at hand, we choose $\bar{g}$ to be given by the line element (\ref{eq:line}) with $Q_I=1$ for $I\leq 4$, $Q_I=0$ for $I\geq6$ and $Q_5 = \Omega$, \ie empty AdS$_4$.

It is easy to show that any solution to $G_{ab}=0$ with $\xi=0$ is a solution to $G^{H}_{ab}=0$. However, the converse is not necessarily true. In certain circumstances one can show that solutions with $\xi\neq 0$, coined Ricci solitons, cannot exist \cite{Figueras:2011va}. There are two key assumptions of the theorems detailed in \cite{Figueras:2011va}: the system of equations is Elliptic, and no matter besides a cosmological constant is included. While the second condition is met in our case, the first is certainly not obvious. In fact, our system of equations appears to be of the mixed Elliptic-Hyperbolic type, for which the De-Turck method seems to work \cite{Fischetti:2012vt,Figueras:2012rb}, even though we do not have a current understanding of why that is the case. In a nutshell, we are going to use Eq.~(\ref{eq:einsteindeturck}), and a posteriori check that each of the components of $\xi$ are zero to machine precision. The advantage of using Eq.~(\ref{eq:einsteindeturck}) instead of the original Einstein equations (\ref{eq:einstein}) is that no explicit gauge condition needs to be imposed prior to our numerical construction. Also, the system of equations Eq.~(\ref{eq:einsteindeturck}) can be shown to be quasi-linear in this gauge, which is a major advantage numerically.

The boundary conditions we use follow from regularity and normalizability at the conformal boundary. At $x=0$ we demand:
\begin{subequations}
\begin{equation}
\left\{
\begin{array}{cc}
Q_I(0,y,\psi)=0& I\in\{7,8,9\}
\\
\\
\partial_x Q_I(x,y,\psi)|_{x=0}=0& \text{otherwise}
\end{array}\right.\,,
\end{equation}
whereas at the north pole $x=1$
\begin{equation}
\left\{
\begin{array}{cc}
Q_I(1,y,\psi)=0& I\in\{7,8\}
\\
\\
\partial_x Q_I(x,y,\psi)|_{x=1}=0& \text{otherwise}
\end{array}\right.\,.
\end{equation}
At the conformal boundary we demand
\begin{equation}
\left\{
\begin{array}{cc}
Q_I(x,1,\psi)=1& I\in\{1,2,3,4\}
\\
\\
Q_5(x,1,\psi)=\Omega&
\\
\\
Q_I(x,1,\psi)=0& \text{otherwise}
\end{array}\right.\,,
\end{equation}
and at the AdS$_4$ center
\begin{equation}
\partial_y Q_I(x,y,\psi)|_{y=0}=0\,.
\end{equation}
\label{eqs:bcs}
\end{subequations}

In order to solve Eq.~(\ref{eq:einsteindeturck}) subject to the boundary conditions (\ref{eqs:bcs}), we use a standard pseudospectral collocation approximation in $y$, $x$ and $\psi$ and solve the resulting non-linear algebraic equations using a damped Newton-Raphson method. We represent the dependence in $y$ and $x$ of all functions as a series in Chebyshev polynomials and the $\psi$-dependence as a Fourier series, so the $\psi$-direction is periodically identified.

Starting with the $\ell_s = m_s =2$ linearized mode as the seed solution in the Newton's method, we obtain a  one parameter family of exact (numerical) solutions based on this mode.  Note that at the non-linear level all higher modes with $\ell_s\geq 2$ are excited, but have exponentially decaying amplitude with increasing $\ell_s$. Since one could start with any mode, we expect a countable infinite number of (one parameter) families of geons labelled by their respective quantum numbers. However, we also expect local uniqueness of solutions, since the system of PDEs we are solving is elliptic in some regions of spacetime.

\begin{figure}[h]
\centering
\subfigure{\label{figs:genprea} (a)\includegraphics[height = 0.27\textheight]{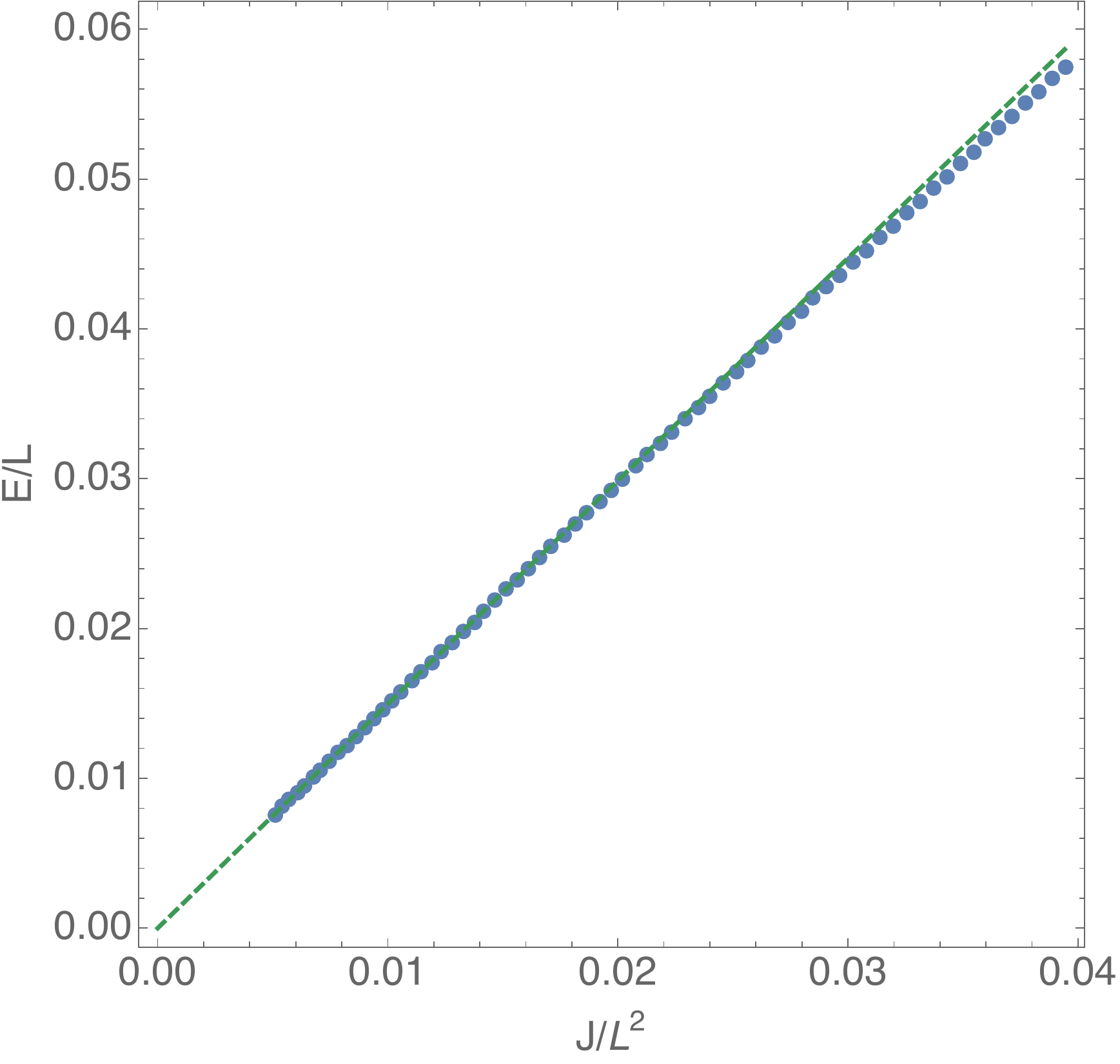}}
\subfigure{\label{figs:genpreb} (b) \includegraphics[height = 0.27\textheight]{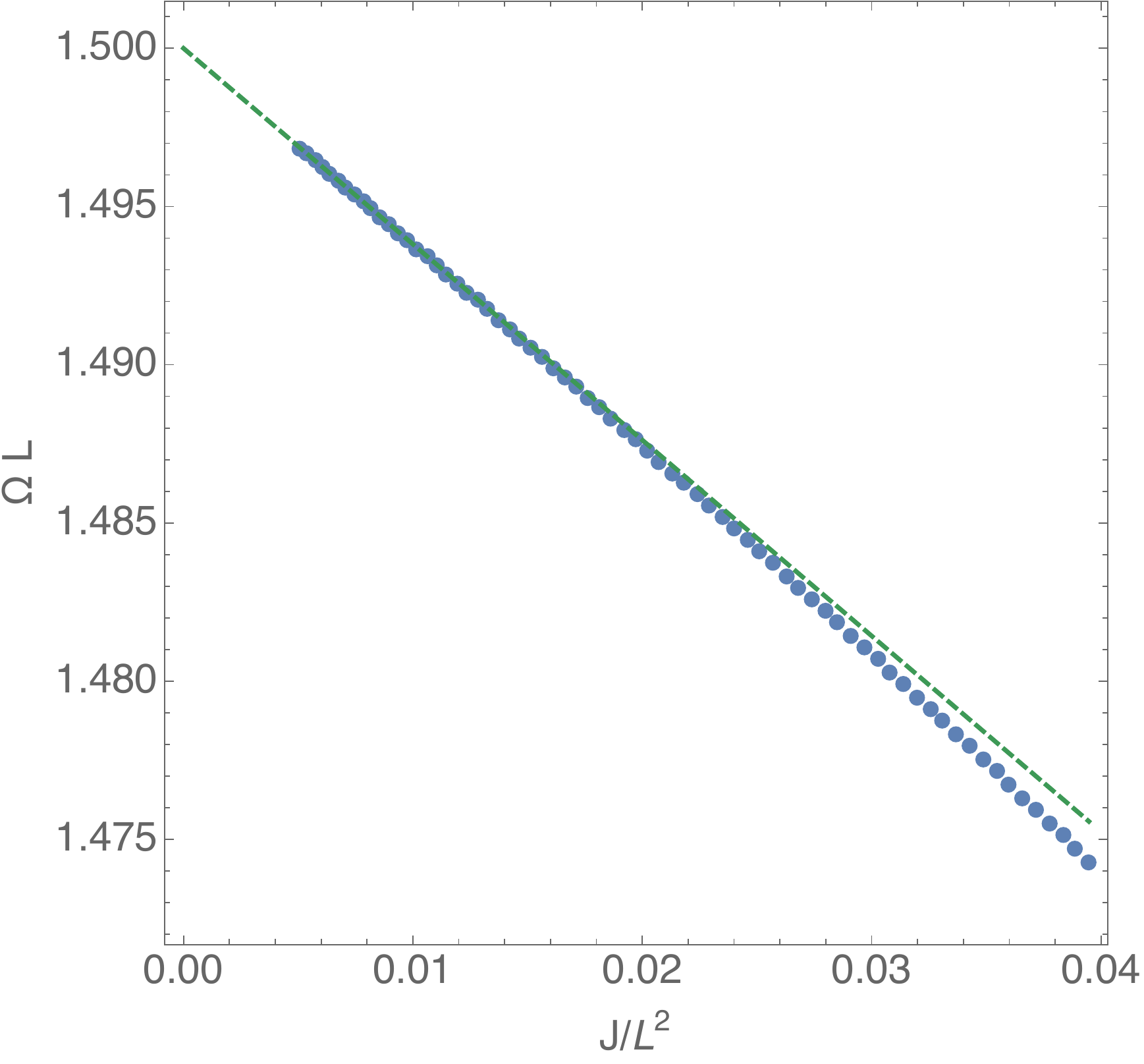}}
\caption{\label{figs:genpre}{ Comparing numerical data with analytic predictions}:  Plot (a) shows the energy of the geon as a function of its angular momentum, while plot (b) shows its angular velocity  as a function of the same quantity. Dashed curves correspond to the analytic predictions \reef{geonEJ} (with $\Omega = \omega_2/2$) while the dots correspond to the numerical data.}
\end{figure}

Due to the helical symmetry, these solutions are exactly time periodic.
They satisfy all of the expected properties for geons. We have computed the total energy \reef{energy} and angular momentum \reef{angmomentum} for these solutions, as well as the angular velocity.
In Fig.~\ref{figs:genprea} we plot the energy as a function of the angular momentum, and in Fig.~\ref{figs:genpreb} the angular velocity as a function of the same quantity. The dashed curves are the perturbative analytic prediction \reef{geonEJ} and the dots are our numerical solutions obtained by solving Eq.~(\ref{eq:einstein}) with the boundary conditions (\ref{eqs:bcs}). It is reassuring that for small enough angular momentum, the analytic predictions match our numerical results.

One can show that any one parameter family of nonsingular solutions to Einstein's equation with symmetry $K $ that approaches $ \partial_t+\Omega\partial_\phi$ asymptotically must satisfy $\delta E={\Omega} \,\delta J$. This is an analog of the first law of black hole mechanics, but now applied to spacetimes without a horizon. It can be derived from the Hamiltonian formulation of general relativity as follows \cite{Sudarsky:1992ty}. As is well known, the Hamiltonian for general relativity takes the form:
\be
H = \int_\Sigma N^a C_a + {\rm surface \ terms}
\ee
where $\Sigma$ is a complete spacelike surface, $N^a$ is the lapse-shift vector, and   $C_a$ are the Hamiltonian and momentum constraints. The surface terms can be determined by the requirement that the variation of $H$ with respect to the canonical variables be well defined, and yields the conserved quantities at infinity.
  Suppose we choose $N^a = K^a$. Then on the one hand, if we perturb the geon, $\delta H = 0$, since the variation of the Hamiltonian gives the evolution of the canonical variables along $N^a$, but this is a symmetry direction. On the other hand, if the perturbation satisfies the linearized field equation, it satisfies the linearized constraints, so $\delta H$ reduces to the variation of the surface terms. This implies $\delta E - {\Omega} \,\delta J = 0$.
We have numerically checked that our extracted energy and angular angular momentum satisfy this to within $10^{-4}\,\%$.

\begin{figure}[h]
\centering
\includegraphics[height = 0.27\textheight]{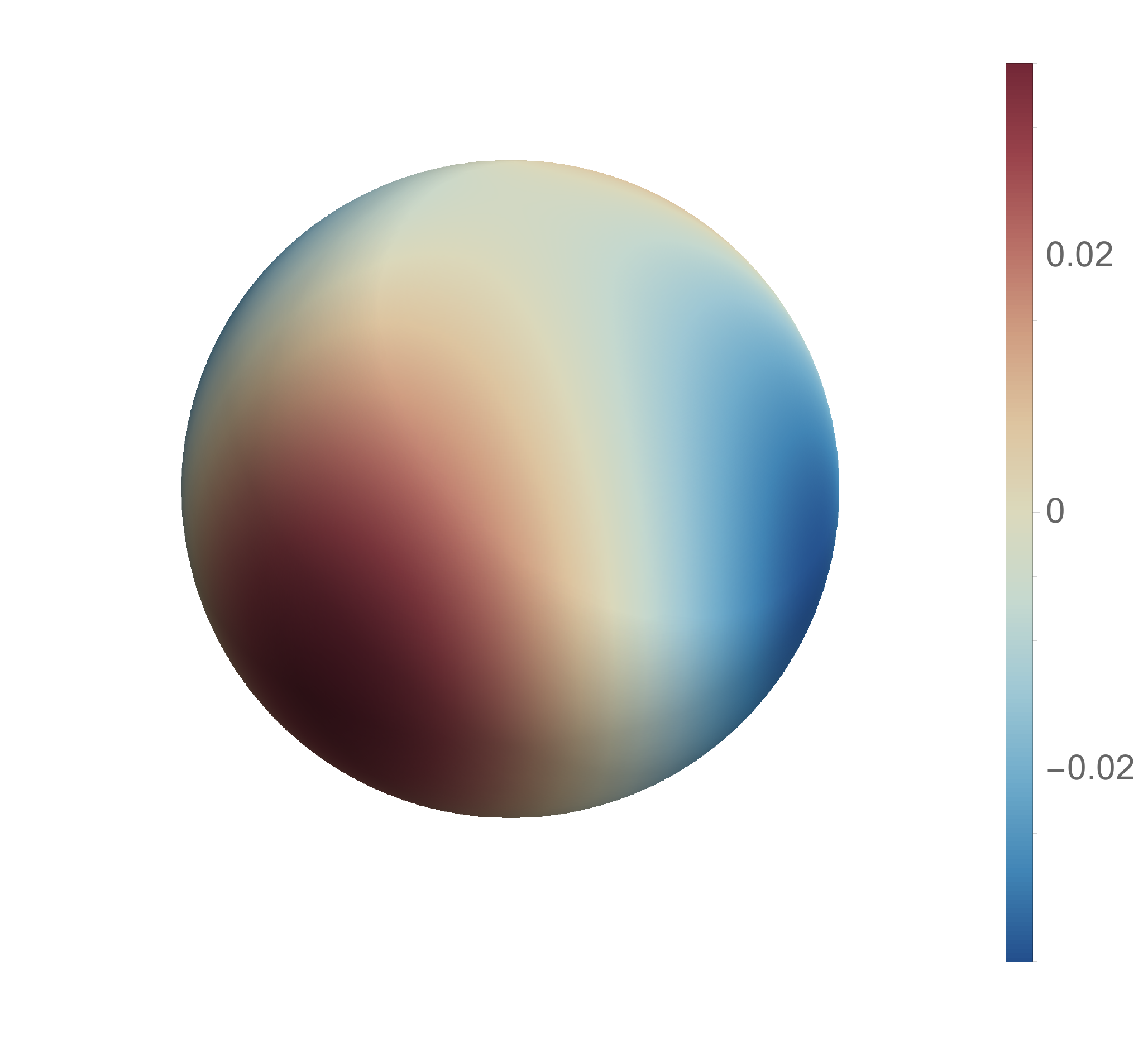}
\caption{\label{fig:stress}{The $G L^2 \langle T_{tt}\rangle$ component of the stress energy tensor for $J/L^2 = 0.02$}: blue regions indicate regions where $G L^2 \langle T_{tt}\rangle$ is negative, while red is positive. Its maximum value is reached at $0.0344$, while its minimum is $-0.0278$.}
\end{figure}

In addition to the energy and angular momentum of the geon, we can also read off its boundary stress energy tensor \reef{bdystressenergy}. It turns out all components are nonzero. For concreteness, we plot in Fig.~\ref{fig:stress} a snapshot of the energy density $G L^2 T_{tt}$ for $J/L^2 = 0.02$. Since the solution has a helical symmetry, the energy density is a function of time, rigidly rotating with constant angular velocity $\Omega$. It contains regions around the equator where its sign changes. Note that while the energy density is negative in certain regions, its integral yields the total energy, which is always positive. The  angular velocity is larger than one in units of the radius of the sphere,\footnote{Recall that $T_{\mu\nu}$ is defined on $S^2 \times R$.} which naively suggests superluminal propagation. However, we note that this is the \emph{phase velocity}, which is not bounded above by causality.

 Finally we comment on the geon's stability. Even linear stability has not been established. 
One  can make arguments both in favor and against its stability. On the one hand, geons are nonlinear realizations of normal modes of AdS, and in particular, when their energy is small, it is natural that their perturbative spectrum is similar to that of AdS. Since AdS does not have any mode that is on the verge of becoming unstable, we expect the geons to be stable\footnote{This is merely an argument, since we cannot prove that the spectrum of linear perturbations around a geon yields a selfadjoint problem, and thus we cannot rule out the existence of an instability.}. On the other hand, the fact that the Killing vector $K $  \reef{defK} is spacelike in the asymptotic region near the equator, indicates the presence of an ergoregion that extends all the way to the conformal boundary. This suggests a possible instability  similar to the one discussed in \cite{Friedman:1978wla}.
 
If geons  are linearly stable, they are likely to be nonlinearly stable as well. This is because there are very few (if any) resonances. There are modes with large angular momentum $\ell$ which are concentrated in the asymptotic region and have frequencies close to AdS, so there are approximate resonances. But one can show \cite{Dias:2012tq} that for large $\ell$, 
\be
\omega - \omega_{AdS} \propto \ell^{-1/2}
\ee
A formal perturbation argument suggests that this difference is large enough to avoid triggering an instability \cite{Dias:2012tq}. This argument applies to all asymptotically AdS solutions such as AdS black holes.


\section{Discussion and open questions}

Although we have focussed on four dimensions, there is nothing special about this choice. We expect similar results for both the existence of geons and nonlinear instability of higher dimensional AdS. 

It seems likely that one can put a small rotating black hole inside the geons. The main constraint is that the angular velocity of the black hole must agree with the angular velocity of the geon so the Killing field of the geon agrees with the Killing field which is null on the horizon. This is suggested by a similar construction for a  rotating scalar field solution in AdS called a rotating boson star. In that case it has been shown that putting a  small rotating black hole in the center does not destabilize the solution \cite{Dias:2011at}. If rotating black holes do exist inside geons, it would clearly show that Kerr-AdS is not the only stationary, asymptotically AdS vacuum black hole.

 We conclude with some open questions raised by this work.  The main one is: 
  
\begin{enumerate} 
\item {\it Is AdS nonlinearly unstable to generic gravitational perturbations?}  As we have discussed, there is a perturbative argument that the answer is yes. 

If so, one can ask about both the origin and endpoint of this instability:

\item {\it Is this instability due to the fact that AdS has a resonant linear spectrum?} There is a perturbative argument that it is, which would imply that other asymptotically AdS solutions should be stable.

\item {\it Do generic nonlinear gravitational perturbations of AdS evolve to form small black holes?} We have seen that this is true for spherical scalar fields coupled to gravity. But since gravitational modes involve angular momentum, it is possible that there will be a barrier to forming small black holes.

 The next two questions regard properties of geons:
 
\item {\it Are geons stable?} As we mentioned above, this question is open even at the linear level. Assuming it is linearly stable, one can ask about nonlinear perturbations. If the answer to (2) is yes, one would expect nonlinear stability.

\item {\it Is there a maximum mass for the geons?} This is expected since there is a maximum mass for the analogous solutions involving scalar fields:  rotating boson stars
\cite{Dias:2011at}.

 Finally one can ask about the stability of AdS black holes:

\item {\it Are AdS black holes nonlinearly stable?} Linear perturbations of black holes are usually described in terms of quasinormal modes with complex frequencies, so individual modes decay exponentially in time. However, it has been shown that  perturbations of AdS black holes with finite Sobelov norm only decay like $1/(\log t)^p$ where the exponent depends on which Sobelov space one starts with \cite{Holzegel:2011uu}. Even analytic perturbations typically fall-off only as a power of $1/t$.  This slow decay might indicate an instability. On the other hand, perturbative arguments based on resonances indicate that AdS black holes should be nonlinearly stable \cite{Dias:2012tq}.

 \end{enumerate}

 \vskip 1cm
\centerline{\bf Acknowledgements}
\vskip 1cm

We wish to thank O. Dias for his collaboration on the work described in section 2. This work was supported in part by NSF grant PHY12-05500. It was also supported in part by NSF grant  PHYS-1066293 and the hospitality of the Aspen Center for Physics.


\bibliographystyle{JHEP}
\bibliography{geons}

\providecommand{\href}[2]{#2}\begingroup\raggedright\begin{thebibliography}{10}

\bibitem{Friedrich:1995vb}
H.~Friedrich, {\it {Einstein equations and conformal structure - Existence of
  anti de Sitter type space-times}},  {\em J.Geom.Phys.} {\bf 17} (1995)
  125--184.

\bibitem{Balasubramanian:1999re}
V.~Balasubramanian and P.~Kraus, {\it {A Stress tensor for Anti-de Sitter
  gravity}},  {\em Commun.Math.Phys.} {\bf 208} (1999) 413--428,
  [\href{http://xxx.lanl.gov/abs/hep-th/9902121}{{\tt hep-th/9902121}}].

\bibitem{deHaro:2000xn}
S.~de~Haro, S.~N. Solodukhin, and K.~Skenderis, {\it {Holographic
  reconstruction of space-time and renormalization in the AdS / CFT
  correspondence}},  {\em Commun.Math.Phys.} {\bf 217} (2001) 595--622,
  [\href{http://xxx.lanl.gov/abs/hep-th/0002230}{{\tt hep-th/0002230}}].

\bibitem{0602.53007}
C.~Fefferman and C.~Graham, ``{Conformal invariants.}.'' {\'Elie Cartan et les
  math\'ematiques d'aujourd'hui, The mathematical heritage of \'Elie Cartan,
  S\'emin. Lyon 1984, Ast\'erisque, No.Hors S\'er. 1985, 95-116}, 1985.

\bibitem{Christodoulou:1993uv}
D.~Christodoulou and S.~Klainerman, {\it {The Global nonlinear stability of the
  Minkowski space}},  {\em Princeton University Press} (1993).

\bibitem{Friedrich86}
H.~Friedrich, {\it {On the existence of n-geodesically complete or future
  complete solutions of EinsteinÕs field equations with smooth asymptotic
  structure}},  {\em Commun. Math. Phys.} {\bf 107} (1986) 587.

\bibitem{Bizon:2011gg}
P.~Bizon and A.~Rostworowski, {\it {On weakly turbulent instability of anti-de
  Sitter space}},  {\em Phys.Rev.Lett.} {\bf 107} (2011) 031102,
  [\href{http://xxx.lanl.gov/abs/1104.3702}{{\tt arXiv:1104.3702}}].

\bibitem{Choptuik:1992jv}
M.~W. Choptuik, {\it {Universality and scaling in gravitational collapse of a
  massless scalar field}},  {\em Phys.Rev.Lett.} {\bf 70} (1993) 9--12.

\bibitem{Christodoulou:1991yfa}
D.~Christodoulou, {\it {The formation of black holes and singularities in
  spherically symmetric gravitational collapse}},  {\em Commun.Pure Appl.Math.}
  {\bf 44} (1991), no.~3 339--373.

\bibitem{Maliborski:2013jca}
M.~Maliborski and A.~Rostworowski, {\it {Time-Periodic Solutions in an Einstein
  AdSÐMassless-Scalar-Field System}},  {\em Phys.Rev.Lett.} {\bf 111} (2013),
  no.~5 051102, [\href{http://xxx.lanl.gov/abs/1303.3186}{{\tt
  arXiv:1303.3186}}].

\bibitem{Dias:2011ss}
O.~J. Dias, G.~T. Horowitz, and J.~E. Santos, {\it {Gravitational Turbulent
  Instability of Anti-de Sitter Space}},  {\em Class.Quant.Grav.} {\bf 29}
  (2012) 194002, [\href{http://xxx.lanl.gov/abs/1109.1825}{{\tt
  arXiv:1109.1825}}].

\bibitem{Regge:1957td}
T.~Regge and J.~A. Wheeler, {\it {Stability of a Schwarzschild singularity}},
  {\em Phys.Rev.} {\bf 108} (1957) 1063--1069.

\bibitem{Edelstein:1970sk}
L.~Edelstein and C.~Vishveshwara, {\it {Differential equations for
  perturbations on the schwarzschild metric}},  {\em Phys.Rev.} {\bf D1} (1970)
  3514--3517.

\bibitem{Vishveshwara:1970cc}
C.~Vishveshwara, {\it {Stability of the schwarzschild metric}},  {\em
  Phys.Rev.} {\bf D1} (1970) 2870--2879.

\bibitem{Zerilli:1970se}
F.~J. Zerilli, {\it {Effective potential for even parity Regge-Wheeler
  gravitational perturbation equations}},  {\em Phys.Rev.Lett.} {\bf 24} (1970)
  737--738.

\bibitem{Zerilli:1971wd}
F.~Zerilli, {\it {Gravitational field of a particle falling in a schwarzschild
  geometry analyzed in tensor harmonics}},  {\em Phys.Rev.} {\bf D2} (1970)
  2141--2160.

\bibitem{Ishibashi:2004wx}
A.~Ishibashi and R.~M. Wald, {\it {Dynamics in nonglobally hyperbolic static
  space-times. 3. Anti-de Sitter space-time}},  {\em Class.Quant.Grav.} {\bf
  21} (2004) 2981--3014, [\href{http://xxx.lanl.gov/abs/hep-th/0402184}{{\tt
  hep-th/0402184}}].

\bibitem{Craps:2014vaa}
B.~Craps, O.~Evnin, and J.~Vanhoof, {\it {Renormalization group, secular term
  resummation and AdS (in)stability}},
  \href{http://xxx.lanl.gov/abs/1407.6273}{{\tt arXiv:1407.6273}}.

\bibitem{Maliborski:2014rma}
M.~Maliborski and A.~Rostworowski, {\it {What drives AdS unstable?}},  {\em
  Phys.Rev.} {\bf D89} (2014) 124006,
  [\href{http://xxx.lanl.gov/abs/1403.5434}{{\tt arXiv:1403.5434}}].

\bibitem{cardoso}
H.~Okawa, V.~Cardoso, and P.~Pani, ``{On the nonlinear instability of confined
  geometries}.'' {In preparation}.

\bibitem{Balasubramanian:2014cja}
V.~Balasubramanian, A.~Buchel, S.~R. Green, L.~Lehner, and S.~L. Liebling, {\it
  {Holographic Thermalization, stability of AdS, and the
  Fermi-Pasta-Ulam-Tsingou paradox}},
  \href{http://xxx.lanl.gov/abs/1403.6471}{{\tt arXiv:1403.6471}}.

\bibitem{Headrick:2009pv}
M.~Headrick, S.~Kitchen, and T.~Wiseman, {\it {A New approach to static
  numerical relativity, and its application to Kaluza-Klein black holes}},
  {\em Class.Quant.Grav.} {\bf 27} (2010) 035002,
  [\href{http://xxx.lanl.gov/abs/0905.1822}{{\tt arXiv:0905.1822}}].

\bibitem{Figueras:2011va}
P.~Figueras, J.~Lucietti, and T.~Wiseman, {\it {Ricci solitons, Ricci flow, and
  strongly coupled CFT in the Schwarzschild Unruh or Boulware vacua}},  {\em
  Class.Quant.Grav.} {\bf 28} (2011) 215018,
  [\href{http://xxx.lanl.gov/abs/1104.4489}{{\tt arXiv:1104.4489}}]. Temporary
  entry.

\bibitem{Fischetti:2012vt}
S.~Fischetti, D.~Marolf, and J.~E. Santos, {\it {AdS flowing black funnels:
  Stationary AdS black holes with non-Killing horizons and heat transport in
  the dual CFT}},  {\em Class.Quant.Grav.} {\bf 30} (2013) 075001,
  [\href{http://xxx.lanl.gov/abs/1212.4820}{{\tt arXiv:1212.4820}}].

\bibitem{Figueras:2012rb}
P.~Figueras and T.~Wiseman, {\it {Stationary holographic plasma quenches and
  numerical methods for non-Killing horizons}},  {\em Phys.Rev.Lett.} {\bf 110}
  (2013) 171602, [\href{http://xxx.lanl.gov/abs/1212.4498}{{\tt
  arXiv:1212.4498}}].

\bibitem{Sudarsky:1992ty}
D.~Sudarsky and R.~M. Wald, {\it {Extrema of mass, stationarity, and staticity,
  and solutions to the Einstein Yang-Mills equations}},  {\em Phys.Rev.} {\bf
  D46} (1992) 1453--1474.

\bibitem{Friedman:1978wla}
J.~L. Friedman, {\it {Generic instability of rotating relativistic stars}},
  {\em Commun.Math.Phys.} {\bf 62} (1978), no.~3 247--278.

\bibitem{Dias:2012tq}
O.~J. Dias, G.~T. Horowitz, D.~Marolf, and J.~E. Santos, {\it {On the Nonlinear
  Stability of Asymptotically Anti-de Sitter Solutions}},  {\em
  Class.Quant.Grav.} {\bf 29} (2012) 235019,
  [\href{http://xxx.lanl.gov/abs/1208.5772}{{\tt arXiv:1208.5772}}].

\bibitem{Dias:2011at}
O.~J. Dias, G.~T. Horowitz, and J.~E. Santos, {\it {Black holes with only one
  Killing field}},  {\em JHEP} {\bf 1107} (2011) 115,
  [\href{http://xxx.lanl.gov/abs/1105.4167}{{\tt arXiv:1105.4167}}].

\bibitem{Holzegel:2011uu}
G.~Holzegel and J.~Smulevici, {\it {Decay properties of Klein-Gordon fields on
  Kerr-AdS spacetimes}},  {\em Commun.Pure Appl.Math.} {\bf 66} (2013)
  1751--1802, [\href{http://xxx.lanl.gov/abs/1110.6794}{{\tt
  arXiv:1110.6794}}].

\end{thebibliography}\endgroup

\end{document}